\documentclass[preprint,plb]{elsarticle}
\usepackage{epsfig,graphicx,float,pst-all}
\usepackage{amssymb}
\newcommand{\be}{\begin{equation}} 
\newcommand{\ee}{\end{equation}}
\newcommand{\bc}{\begin{center}}
\newcommand{\ec}{\end{center}}

\renewcommand{\vec}[1]{\mbox{\boldmath $#1$}}

\begin{document}
\begin{frontmatter}

\title{Algebraic model for single-particle energies of 
$\Lambda$ hypernuclei}

\author{L. Fortunato}
\ead{fortunat@pd.infn.it}
\address{
Dipartimento di Fisica e Astronomia ``G.Galilei'' - Universit\`a di Padova, Italy}
\address{
I.N.F.N. - Sez. di Padova, via Marzolo,8, Padova, I-35131, Italy}

\author{K. Hagino}
\ead{hagino@nucl.phys.tohoku.ac.jp}
\address{
Department of Physics, Tohoku University, Sendai 980-8578,  Japan}
\address{Research Center for Electron Photon Science, Tohoku
University, 1-2-1 Mikamine, Sendai 982-0826, Japan}
\address{
National Astronomical Observatory of Japan 2-21-1 Osawa,
Mitaka, Tokyo 181-8588, Japan}

\begin{abstract}
A model is proposed for the spectrum of $\Lambda$ hypernuclei 
based on the  $u(3)\times u(2)$ Lie algebra, in which 
the internal degrees of freedom of the spin-1/2 $\Lambda$ 
particle are treated 
in the Fermionic $u(2)$ scheme,  
while the motion of the hyperon inside a nucleus is described 
in the Bosonic $u(3)$ 
harmonic oscillator scheme.
Within this model, a simple formula for single-particle 
energies of the $\Lambda$ particle is obtained from 
the natural dynamical 
symmetry. The formula is applied to 
the experimental data on the reaction spectroscopy for 
the $^{89}_\Lambda$Y and $^{51}_\Lambda$V hypernuclei, 
providing a clear theoretical interpretation of the observed structures. 
\end{abstract}

\begin{keyword}
Hypernuclei \sep Lambda \sep Algebraic models
\PACS 21.80.+a \sep 21.10.-k \sep 21.60.Fw
\end{keyword}

\end{frontmatter}

\section{Introduction}

Hypernuclear physics has rapidly been developed in the past 
few decades \cite{Hashi}, deepening our knowledge of how hyperons, 
{\it i.e.}, particles containing at least one $s$ quark, 
interact with nuclear matter. 
A characteristic feature of hyperons is that they 
are free from the Pauli blocking due to the nucleons, 
and thus they can probe the deep interior of atomic nuclei. 
Furthermore, they may attract surrounding nucleons, leading to 
important modifications in the nuclear structure 
\cite{Tanida01,Motoba83,Hiyama96,Hiyama99,Myaing08,Isaka11}. 

One of the most remarkable findings in hypernuclear 
physics is the spectral signature of a clear single-particle structure of 
$\Lambda$ particle in single-$\Lambda$ 
hypernuclei \cite{Hashi,Hot,Nakamura13,Gogami16}. 
In particular, single-particle $\Lambda$ states with orbital angular 
momentum ranging from $l=0$ to $l=3$ have been 
clearly identified in medium-heavy $^{89}_\Lambda$Y hypernucleus using the 
$(\pi^+,K^+)$ reaction spectroscopy \cite{Hashi,Hot}. 
These single-particle levels have been theoretically analyzed with the 
distorted-wave impulse approximation (DWIA) based on shell 
model calculations \cite{Motoba08}.

In this paper, we 
re-analyze the experimental data for the 
$(\pi^+,K^+)$ reaction by introducing 
an algebraic model to describe 
single-particle levels of single-$\Lambda$ hypernuclei. 
In this approach, single-particles levels are 
classified according to the underlying symmetries.  
The energy of each level is then given in terms of 
expectation values of Casimir operators associated with the 
dynamical symmetries. 
An advantage of the algebraic method is that 
the spectrum can be predicted 
with a minimal set of requirements associated with symmetry 
even when 
the exact shape of mean-field potential experienced 
by the hyperon is not known. 
In this sense, the model is applicable 
for a whole class of potentials that 
share the same asymptotic behavior.
In fact, the formalism which we present in this paper 
is general enough, and we expect it to be universally valid, 
even beyond the applications 
presented in this paper.

We mention that a mass formula for $\Lambda$ hypernuclei has been 
constructed in Ref. \cite{LCIJ98} based on a similar algebraic approach. 
In contrast to Ref. \cite{LCIJ98}, 
our interest in this paper is 
in the excitation spectra of 
single-$\Lambda$ hypernuclei, rather than 
the ground state mass. 

\section{Algebraic model for hypernuclei}

Our aim in this paper is to describe the spectrum of hypernuclei 
with an algebraic model. To this end, 
one can model the total Hamiltonian in two terms: 
\be
\hat H= \hat H_{\rm Nucl}+ \hat H_{\rm Hyp}, 
\label{ham}
\ee
where $\hat H_{\rm Nucl}$
is for the core nucleus without hyperons, and 
$\hat H_{\rm Hyp}$ for the hyperons interacting with 
non-strange nucleons. 
In this paper, we are not concerned with the specific model 
for the core nucleus, $\hat H_{\rm Nucl}$: 
it might be {\it e.g.,} a shell-model, 
an interacting boson model (IBM) based on U(6) symmetry, 
a collective model, or any other suitable model, 
provided that it gives eigenstates with good total angular 
momentum $J$. As a matter of fact, for simplicity, 
we assume that the core nucleus remains in the ground state with $J=0$, 
and discard $\hat H_{\rm Nucl}$ from Eq. (\ref{ham}) 
in the following discussion. 

We model the hyperon part of the Hamiltonian, $\hat H_{\rm Hyp}$, as 
follows. First we notice that, despite that the nuclear medium is so dense, 
the quantized motion of the hyperon through it proceeds 
with an almost constant attractive interaction within the nuclear volume, 
because there is no Pauli blocking effect. 
The $\Lambda$ particle then bounces back at the surface, 
being attracted inward by a restoring force. 
The bosonic algebra of the three-dimensional harmonic oscillator 
$u(3)$ can be used to model this situation, in which 
the motion of any other strange or charmed chargeless 
particle would behave similarly.
We then couple 
a $u_F(2)$ fermionic Lie algebra scheme for $\Lambda$ spin-1/2 particles 
to the bosonic $u_B(3)$ Lie algebra for the harmonic motion of 
the hyperon in the nuclear interior. 

The only possible dynamical symmetry arising in this simple scheme 
is given by the 
\be
u(3/2) \supset u_B(3)\times u_F(2) \supset so_B(3)\times su_F(2)
\ee
subalgebra chain, where for our purposes we can discard the 
lower $so(2)$ symmetries as they will affect only magnetic substates. 
In this paper, we consider the following 
model Hamiltonian:
\be
\hat H_{\rm Hyp} = \hat H_{\rm u(3)} + \hat H_{\rm u(2)} + \hat V_{\rm int},
\label{Hhyp}
\ee
where the $u(3)$ part provide 
a global mean-field dynamics and 
the second term, $\hat H_{\rm u(2)}$, 
is a constant energy that depends on the number of 
$\Lambda$ particle ($n$ = 0, 1, and 2). 
The last term, $\hat V_{\rm int}$, describes 
any additional coupling interaction such as a spin-orbit coupling, 
that we will discuss later. 
For simplicity, we will neglect higher order terms in the present analysis.

\subsection{$u(3)$ harmonic oscillator}

Let us first discuss the $u(3)$ part in Eq. (\ref{Hhyp}). 
Up to two body, the simplest and yet already general Hamiltonian with 
dynamical symmetry reads:
\be
\hat H_{\rm u(3)}= \alpha\, \hat C_1(u(3)) +\beta\, \hat C_2(u(3))
+ \gamma\, \hat C_2(so_L(3))
\ee
where $\alpha, \beta$, and $\gamma$ are free parameters and 
the Casimir operators are given by 
$\hat C_1(u(3))=\hat N$, $\hat C_2(u(3))=\hat {N^2}$ and 
$C_2(so_L(3))=\hat {\vec{L}^2}$. 
The spectrum of this Hamiltonian is given in term of the eigenvalues 
of the Casimir operators by 
\be
E_{\rm u(3)}=  \alpha N+  \beta N(N+2)+ \gamma L(L+1)
\label{hospec}
\ee
where $N$ is the number of quanta and $L$ is the orbital 
angular momentum of the confined $\Lambda$ particle inside the nuclear volume.
This clearly gives an (an)harmonic spectrum with rotational-vibrational levels.
The allowed symmetric representations of $u(3)$ are 
labeled by integers $N=0,1,2,\cdots$ and, for each value of $N$,  
possible values of $L$ are given by 
\be
L = N,N - 2, \dots ,1 \mbox{ or } 0 \qquad (N = \mbox{ odd or even })
\label{rule}
\ee
(see Sect.7.5.1 in Ref. \cite{Iac}). 

\subsection{$u(2)$ algebra for $\Lambda$ fermion with $s=1/2$}

Let us next consider the $u(2)$ part. 
We associate a fermionic creation and an annihilation operators 
to each substate of the $s={1/2}$ state as, 
$$ a^\dag_{1/2,+1/2} \qquad a^\dag_{1/2,-1/2}$$
such that their anticommutator reads
\be
\Bigl\{ a_{1/2,m},a^\dag_{1/2,m'} \Bigr\} =  \delta_{m,m'}  \;.
\ee
With the bilinear products of $a$ and $a^\dagger$, 
one can construct the $u(2)$ Lie algebra. 
The four elements for this algebra are 
the total spin operator, $\hat S_\mu$, 
with $\mu=0,\pm1$ and 
the number operator for fermions, 
$\hat N_F$ (see Sect. 8.4.2 in Ref. \cite{Iac}).
These 4 elements are related to 
a vector operator defined as 
\be
A^{(1)}_\mu =\Bigl[ a^\dag_{1/2} \times {\tilde a}_{1/2}\Bigr]^{(1)}_\mu 
= -\sqrt{2} \hat S_\mu, 
\ee
and a scalar operator defined as 
\be
A^{(0)}_0 =\Bigl[ a^\dag_{1/2} \times {\tilde a}_{1/2}\Bigr]^{(0)}_0 
= -\sqrt{\frac{1}{2}} \hat N_F. 
\ee
Here, we have used 
the definition ${\tilde a}_{j,m} = (-1)^{j-m} a_{j,-m}$. In our case with 
$s=1/2$, ${\tilde a}_{1/2,+1/2}= a_{1/2,-1/2}$ and 
${\tilde a}_{1/2,-1/2}= -a_{1/2,+1/2}$.

The relevant linear combination of Casimir operators 
for the fermionic part of the Hamiltonian is: 
\be
\hat H_{\rm u(2)} =A \hat C_1\bigl(u(2)\bigr) + B\hat C_2\bigl(u(2)\bigr),   
\ee
and the corresponding energy formula reads, 
\be
E_{\rm u(2)} = A \langle C_1\rangle  + B\langle C_2\rangle. 
\ee
The representations of $u(2)$ are given in general 
by a pair of numbers $[\lambda_1, \lambda_2]$. 
Using the algorithm in 
Sect. 5.4.1 in 
Ref. \cite{Iac},  
one can calculate the eigenvalues of the linear and quadratic Casimir 
operators as: 
\be
\langle C_1\rangle = \lambda_1+\lambda_2,
\qquad
\langle C_2\rangle = \lambda_1^2+\lambda_1+\lambda_2^2-\lambda_2.
\ee
Their eigenvalues in fermionic (antisymmetric) representations are 
thus given by, 
\be
\begin{tabular}{lccc}
  & [0] & [1] & [1,1] \\
$\langle C_1\rangle$ & 0 & 1& 2 \\
$\langle C_2\rangle$ & 0 & 2& 2 \\
\end{tabular}
\ee
where the notation $[0]$, $[1]$ and $[1,1]$ means zero fermions, 
one fermion (in either spin state) and two fermions, respectively. 

Since there are only three possible fermionic states, 
the formula can take the values of 
\be
E_{0} =0, \qquad
E_{\Lambda} = A+2B, \qquad
E_{\Lambda\Lambda} = 2A+2B.  
\label{syst}
\ee
Together with 
Eq. (\ref{hospec}), the energies of hypernuclei then read
\be
E_{\rm Hyp} = 
\alpha N+  \beta N(N+2)+ \gamma L(L+1) +E_{n\Lambda},
\label{Ehyp}
\ee
where the last term is given by Eq. (\ref{syst}), depending on 
the number of $\Lambda$ particles in the system. 

\section{Reanalysis of the experimental data}

We now apply the energy formula introduced in the previous section to 
single-$\Lambda$ hypernuclei and reanalyze 
the experimental data 
obtained by Hotchi {\it et al.} \cite{Hot} for $^{89}_\Lambda$Y 
and $^{51}_\Lambda$V hypernuclei. 
The measured cross-sections (integrated in the 2$^o$-14$^o$ range) 
for the ($\pi^+, K^+$) reaction leading to the formation 
of the hypernuclei 
show several peaks as a function of energy \cite{Hot}, 
that are interpreted 
as corresponding to different 
angular momentum states of the $\Lambda$ particle. 

In Ref. \cite{Hot}, the experimental data for $^{89}_\Lambda$Y 
have been empirically 
fitted with a combination of 10 Gaussian functions 
(with a total of 18 parameters, 8 of which are energy centroids 
and the remaining 10 are connected to the height of each peak). 
With this procedure, the authors of Ref. \cite{Hot} have concluded 
that the observed broad bumps contain at 
least two sub-peaks. 
The width of the bumps in the spectrum has been attributed to i) 
the experimental energy resolution, that was estimated to be 1.65 MeV 
for this hypernucleus \cite{Hot} and ii) the spreading width 
due to presence of several low-lying 
excited states of the core nucleus. 

The fit obtained in Ref. \cite{Hot} well reproduces the experimental 
spectra, but it lacks a theoretical understanding, 
even though 
it would be essentially correct that the peaks are associated 
with growing angular momenta of $\Lambda$ particle. 
We therefore re-fit here each major peak 
using 
a mathematically complete formalism with all quantum 
numbers attributed to the $u(3)\times u(2)$ chain. 
That is, each major peak is reassigned 
to the different harmonic oscillator shells 
with increasing $N$, whereas 
the lower component within each peak is 
assigned to the largest possible $L$ 
according to the rule given by Eq. (\ref{rule}). 
For example, the first peak has $N=0$ and therefore only $L=0$, 
while the third peak with $N=2$ has $L=0$ and 2 components. 

To this end, we have 
undertaken a new fit with a Gaussian function given by 
\be
G(E;b_{N,L},\sigma) = 
\frac{1}{\sqrt{2\pi\sigma^2}} e^{\{-(E-b_{N,L})^2/2\sigma^2\}} \;,
\ee
where $b_{N,L}$ is the centroid energy given by 
Eq. (\ref{Ehyp}) with $E_{n\Lambda} = E_\Lambda$ (see Eq. (\ref{syst})). 
We superpose 8 Gaussian functions as, 
\be
G(E)=
\Delta E_{\rm bin}\,\sum_{N=0}^4\sum_L a_{N,L}G(E;b_{N,L},\sigma),
\label{gauss-fit}
\ee
where 
$\Delta E_{\rm bin}$ 
is the bin width.  
The value of $L$ is determined for each $N$ according to the rule, 
Eq. (\ref{rule}), except for $N=4$, for which 
we have found that 
the $L$ = 0 component provides only a negligibly small 
contribution, at least in the energy region where the experimental 
data were taken. 
Following Ref. \cite{Hot}, we have used 
$\Delta E_{\rm bin}$ = 0.25 MeV and $\sigma$ = 1.65 MeV. 
In this way, the fit contains 
12 parameters in total, 8 of which are for heights, $a_{N,L}$, 
and the remaining 4 parameters are for the energy formula, 
Eq. (\ref{Ehyp}). 

\begin{figure}[!t]
\includegraphics[width=0.8\textwidth, clip=, bb= 20 38 715 529]{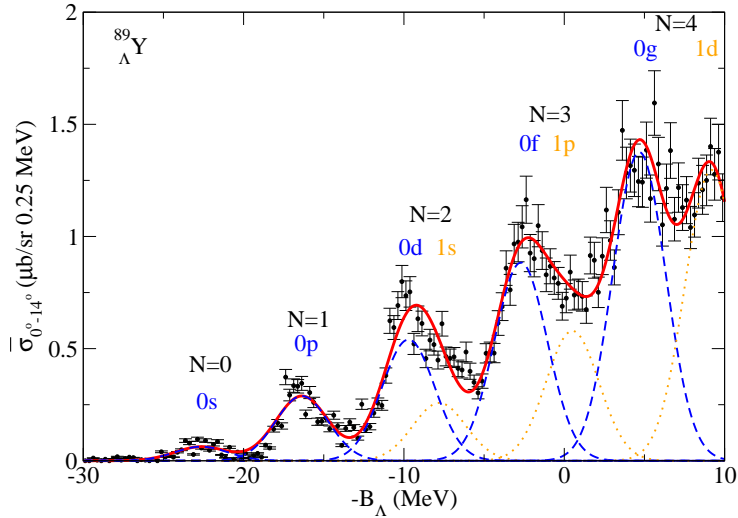}
\caption{
A fit to the experimental 
hypernuclear mass spectra of $^{89}_\Lambda$Y 
based on the algebraic model. 
The dashed and the dotted lines denote each component of the 
major peaks. The experimental data are taken from 
Ref. \cite{Hot}.} 
\label{fit0}
\end{figure}

The resultant fit for the $^{89}_\Lambda$Y hypernucleus 
is shown in Fig. \ref{fit0}, together 
with the parameters in 
Table \ref{par89Y}. 
The quantum number assignments to each Gaussian are also 
shown in the figure, where the components with $n=0$ and 1, 
defined as $N=2n+L$, are shown by the dashed and the dotted lines, 
respectively. 
One can see that the fit is as good as the previous empirical fit. 
While we do not want to put stress on the statistical comparison 
of the fitting procedures or on the fact that we use only 12 parameters 
with respect to 18, we consider that 
this is a considerable gain in the theoretical interpretation of 
the experimental data, because one can now assign quantum numbers 
and determine the splitting of some of the sub-peaks 
on the basis of the algebraic theory. 

\begin{table}[!b]
\begin{center}
\begin{tabular}{cc|cc|cc|cc}
\hline
\hline
$a_{00}$ & 1.03458 & $a_{11}$ & 4.774  & $a_{20}$ & 4.26848 & $a_{22}$ & 8.93652\\
$a_{31}$ & 9.57975 & $a_{33}$ & 14.6199 & $a_{42}$ & 21.4563 & $a_{44}$ & 22.7715\\
\hline
$\alpha$ & 5.39547 & $\beta$  &0.506972 & $\gamma$ & $-$0.321663 
& $E_\Lambda$ & $-$22.6373 \\
\hline
\hline
\end{tabular} 
\caption{The parameters in Eq. (\ref{gauss-fit}) 
(see also Eq. (\ref{Ehyp}))
for the best fit of the 
empirical mass spectra of 
$^{89}_\Lambda$Y. The parameters $\alpha, \beta, \gamma$ and $E_\Lambda$ are in MeV, while $a_{NL}$ are in $\mu$b MeV.}
\label{par89Y}
\end{center}
\end{table}

The fit also shows two other interesting facts. Firstly, $\beta$ is about 
one tenth of $\alpha$, that is, the anharmonicity is indeed small 
but non-negligible for the first few states. 
Secondly, $\gamma$, the coefficient of the $\vec L^2$ term, is negative, 
which implies that the state with higher $L$ comes lower and 
it is usually stronger. 
This confirms that the intuition in Ref. \cite{Hot} 
of assigning the main peaks 
to increasing values of $L$ 
was indeed correct.
At the same time it gives a natural explanation for most of the 
observed features.

\begin{figure}[!t]
\includegraphics[width=0.8\textwidth, clip=, bb= 16 30 579 429]{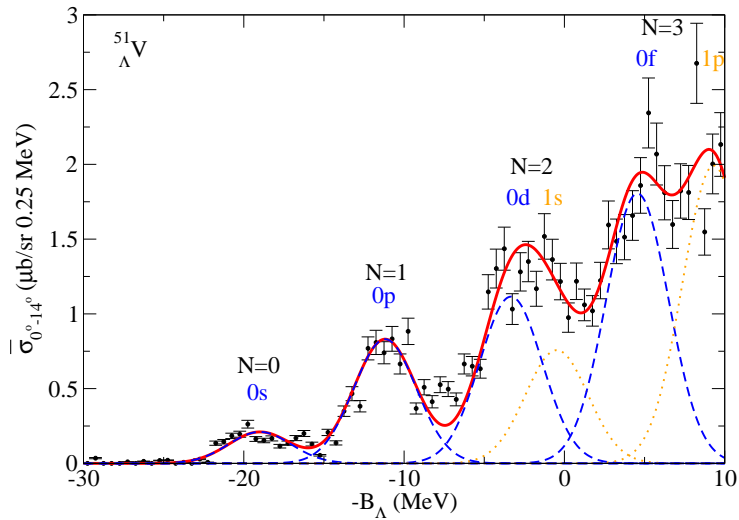}
\caption{Same as Fig. \ref{fit0}, but for the $^{51}_\Lambda$V hypernucleus.}
\label{V-fit0}
\end{figure}

We have repeated 
a similar analysis for the $^{51}_\Lambda$V hypernucleus, and 
have again achieved a good agreement with the experimental data, 
as shown in Fig. \ref{V-fit0}. 
Since both the experimental error bars and the energy resolution 
are larger for this hypernucleus, 
the fit is less accurate as compared to that for $^{89}_\Lambda$Y 
shown in Fig. \ref{fit0}. 
Nevertheless, the present algebraic model predicts 
six peaks in the mass spectra with four major peaks with 
$N=0,\dots, 3$.  
The number of parameters which we employ is 10, 
in which 4 parameters are for the energy formula as before and 
6 for the peak heights (see Table \ref{par}). 

Higher precision experimental data would 
help constraining even more the parameters of the fit. 
It should be noticed, however, 
that 
the energy resolution of the detection apparatus may not be the 
only origin for 
the width of the peaks, 
that is, the excitations of the core nucleus may also contribute to 
the width. 
In this case, 
one would have to perform 
the fitting procedures by taking this effect into account, 
as the true lowest state 
will be found at the lower end (left in the pictures) of each peak. 

\begin{table}[!t]
\begin{center}
\begin{tabular}{cc|cc|cc|cc}
\hline
\hline
$a_{00}$ & 4.11173 & $a_{11}$ & 16.2407  & $a_{20}$ & 14.8101 & $a_{22}$ & 21.8096\\
$a_{31}$ & 38.9592 & $a_{33}$ & 35.3016 &  & \\
\hline
$\alpha$ & 7.2807 & $\beta$  &0.495241 & $\gamma$ & $-$0.476428 
& $E_\Lambda$ & $-$19.003 \\
\hline
\hline
\end{tabular} 
\caption{Same as Table. \ref{par89Y}, but for the $^{51}_\Lambda$V hypernucleus.}
\label{par}
\end{center}
\end{table}

\section{Higher order interactions}

The energy resolution of the experiment reported in Ref. \cite{Hot}
was sufficiently good 
to appreciate the bumps corresponding to each major shell $N$ and 
also to some extent the splitting of the states with different $L$ 
within a given $N$. 
On the other hand, 
if one looks at finer details, some discrepancies can be seen, 
that sometimes exceeds a confidence level of 90\%. 
A further insight of the fine structure can be gained from 
the algebraic model, because the operators that form the $u(2)$ 
and $u(3)$ algebras naturally provide a way to classify 
higher order interactions and perturbations starting from the 
two-body level. 
For example, the simplest 
interaction term is the scalar operator 
obtained by coupling the angular momenta of the two algebras, that is, 
a spin-orbit operator of the form:
\be
V_{int}\propto \hat {\vec L} \cdot \hat {\vec S}, 
\ee 
that gives an energy splitting 
into two components proportional to $2L+1$ 
for each $L\ne 0$ state. 
With this interaction, the 0p state, for example, will separate 
into a $J=3/2$ and a $J=1/2$ peaks.

One should always remember, however, 
that, at these energies, the core nucleus 
may also get excited in the reaction process \cite{Motoba08} (see also the 
discussion in the previous section) and therefore there is an even finer 
structure in each peak that cannot be presently resolved. 
For this reason, one would obtain 
a too high value (about 1-2 MeV) of the spin-orbit splitting if one 
tried to 
obtain information on the magnitude directly from the fit. 
Notice that shell-model \cite{Motoba08,Gal,Millener} 
as well as other experiments on lighter hypernuclei \cite{Hashi}
indicate that the spin-orbit splitting is much smaller, 
of the order of 0.05-0.2 MeV.
It would be an interesting future study to investigate 
how the angular momentum of the nuclear 
excited states and that of the $\Lambda$ particle 
give rise to higher-order 
interactions in the context of the algebraic model. 

\section{Conclusions}

We have introduced a simple algebraic model that accounts 
for the major features observed in spectra of 
a $\Lambda$ particle in medium-heavy nuclei. 
This has allowed us to re-fit the experimental data of the $(\pi^+,K^+)$ 
reaction 
with a theoretical model in which the quantum numbers in each state 
are arranged into an energy formula according to symmetries. 
We have achieved 
a good agreement with the measured spectra, 
to within the limitations of the experimental energy resolution. 

We expect that this algebraic model is universally applicable 
to describe the states of hyperons and other hadrons in the 
nuclear medium. The model also provides a way to classify 
higher order interaction terms and would become useful when 
a finer experimental energy resolution will eventually be attained.

\section*{Acknowledgments}
We thank A.B. Balantekin for useful discussions. 
This work is a part of a larger theoretical campaign aimed at 
{\it Interdisciplinary applications of nuclear theory} 
under the project {\it IN:Theory} of the Univ. of Padova.
L.F. acknowledges financial support within the PRAT~2015 project 
CPDA154713 and {\it Iniziative Coop. Univ.} of the Univ. of Padova.
This work was supported also by JSPS KAKENHI Grant Number 2640263.

\end{document}